\documentclass[12pt,fleqn]{article}
\usepackage{amstext,amsopn}

\DeclareMathOperator{\tr}{tr}

\newcommand*\la{\langle}
\newcommand*\ra{\rangle}
\newcommand*\eq{\begin{equation}}
\newcommand*\en{\end{equation}}
\newcommand*\eqa{\begin{eqnarray}}
\newcommand*\ena{\end{eqnarray}}
\newcommand*\nn{\\ \nonumber}
\newcommand*\ot{\otimes}
\newcommand*\op{\oplus}
\newcommand*\al{\alpha}
\newcommand*\be{\beta}

  \newcommand*\tit[5]{{\em #5}, {#1} {\bf #2}, #3 (#4).}

\newcommand*\prl{Phys. Rev. Lett.}

\newcommand*\jmp{J. Math. Phys.}
\newcommand*\jsp{J. Stat. Phys.}

\begin{document}
\begin{titlepage}
\rightline{October 22, 1999}
\rightline{math-ph/9910037}
\vspace{2em}
\begin{center}{\bf\Large Singlets and reflection symmetric spin systems}\\[2em]
%
%
Elliott H.\ Lieb${}^*$ and Peter Schupp${}^{**}$\\[2em]
{\sl Department of Physics, Princeton University\\
Princeton, New Jersey 08544, USA}\\[6em]
\end{center}
\begin{abstract}
We rigorously establish some exact properties of reflection symmetric
spin systems with antiferromagnetic crossing bonds: 
At least one ground state has total spin zero
and a positive semidefinite coefficient matrix. 
The crossing bonds obey an ice rule.
This augments some previous results which were limited to bipartite spin systems 
and is of particular interest for frustrated spin systems.
\end{abstract}

\vfill
\noindent \hrule
\vskip.2cm
\noindent{{\tiny
\copyright 1999 by the authors. 
Reproduction of this article, in its entirety, by any
means, is permitted for non-commercial purposes.}\\
{\scriptsize E-mail: \ lieb@math.princeton.edu, \
schupp@theorie.physik.uni-muenchen.de\\
${}^*$Work partially supported by US National Science Foundation grant
PHY 98 20650.\\
${}^{**}$Present address:
Sektion Physik der Universit\"at M\"unchen,
Theresienstr.\ 37, 80333 M\"unchen, Germany}}
\end{titlepage}
\newpage
\setcounter{page}{1}

\section{Introduction}

Total spin is often a useful quantum number to classify energy eigenstates of spin
systems. An example is the antiferromagnetic Heisenberg Hamiltonian on a
bipartite lattice, whose energy levels plotted versus total spin form towers of
states. The spin-zero tower extends furthest down the energy scale, the
spin-one tower has the next higher base, and so on, all the way up the spin
ladder: $E(S+1) > E(S)$, where $E(S)$ denotes the lowest energy eigenvalue for
total spin $S$ \cite{LM}. The ground state, in particular, has total
spin zero; it is a singlet. This fact had been suspected for a long time, but
the first rigorous proof was probably given by Marshall~\cite{M} for a 
one-dimensional antiferromagnetic chain with an even number of sites, each with
intrinsic spin-1/2 and with periodic boundary conditions. 
This system is bipartite, it can be split into two
subsystems, each of which contains only every other site, so that all
antiferromagnet bonds are \emph{between} these subsystems.
Marshall bases his proof on a theorem
that he attributes to Peierls: Any ground state of the system, expanded
in terms of $S^{(3)}$-eigenstates has coefficients with alternating signs that
depend on the $S^{(3)}$-eigenvalue of one of the subsystems. After a 
canonical transformation, consisting of a rotation of one of the subsystems
by $\pi$ around the 2-axis in spin space, the theorem simply states that all
coefficients of a ground state can be chosen to be positive. To show that this
implies zero total spin, Marshall
works in a subspace with $S^{(3)}$-eigenvalue $M=0$ and uses translation
invariance. His argument easily generalizes to higher dimensions
and higher intrinsic spin. Lieb, Schultz and Mattis~\cite{LSM}
point out that translational invariance is not really necessary,
only reflection symmetry is needed to relate the two subsystems, and the
ground state is unique in the connected case.
Lieb and Mattis~\cite{LM} ultimately remove the requirement of translation invariance
or reflection symmetry and apply the $M$-subspace method to classify excited
states. Like Peierls they use a Perron-Frobenius type argument to prove that in
the $S^{(3)}$-basis the ground state wave function for the connected case is a
positive vector and it is unique. Comparing this wave function with the 
positive wave function of a simple soluble model in an appropriate
$M$-subspace~\cite{L} they conclude that the ground state has total spin
$S = |S_A - S_B|$, where $S_A$ and $S_B$ are the maximum possible spins
of the two subsystems. (In the antiferromagnetic case $S_A = S_B$ and the ground
state has total spin zero.)
In the present article we reintroduce
reflection symmetry, but for other reasons: we want to exploit methods and
ideas of ``reflection positivity" (see \cite{DLS} and references therein.)
We do not require bipartiteness. 
The main application is to frustrated spin systems similar to the pyrochlore
lattices discussed in \cite{LS}.

\section{Reflection symmetric spin system}

We would like to  consider a spin system that consists of 
two subsystems that are mirror images of one another, except for a rotation by
$\pi$ around the \mbox{2-axis} in spin-space, 
and that has antiferromagnetic crossing bonds 
between corresponding sets of sites of the two subsystems. 
The spin Hamiltonian is
\eq
H = H_L + H_R + H_C ,
\en
and it acts on a tensor product of two identical copies of a Hilbert space that 
carries a representation of SU(2). ``$H_L = \tilde H_R$'' in the
sense that $H_L = h \ot 1$ and
$H_R = 1 \ot \tilde h$, where the tilde shall henceforth denote the rotation
by $\pi$ around the 2-axis in spin-space. We make no further assumptions
about the nature of $H_L$ and $H_R$, in particular we do \emph{not} assume that
these subsystems are antiferromagnetic.
The crossing bonds are of anti-ferromagnetic type in the sense that
$H_C = \sum_A \vec S_A \cdot \vec S_{A'}$, with
$\vec S_A = \sum_{i \in A} j_i \,
\vec s_i$ and $\vec S_{A'} = \sum_{i' \in A'} j_i \, \vec s_{i'}$, where
$A$ is a set of sites in the left subsystem, $A'$ is the corresponding
set of sites in the right subsystem, and $j_i$ are \emph{real} coefficients. 
The intrinsic spins $s_i$ are arbitrary
and can vary from site to site, as long as the whole system is reflection symmetric.
We shall state explicitly when we make further assumptions, e.g., that
the whole system is invariant under spin-rotations.
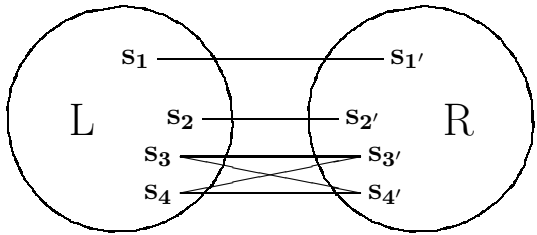
\begin{figure}[h]
\begin{center}
\unitlength 1.00mm
\linethickness{0.4pt}
\begin{picture}(75.00,30.00)
\multiput(20.00,30.00)(0.99,-0.10){3}{\line(1,0){0.99}}
\multiput(22.97,29.70)(0.36,-0.11){8}{\line(1,0){0.36}}
\multiput(25.83,28.82)(0.22,-0.12){12}{\line(1,0){0.22}}
\multiput(28.45,27.39)(0.13,-0.11){17}{\line(1,0){0.13}}
\multiput(30.74,25.47)(0.12,-0.15){16}{\line(0,-1){0.15}}
\multiput(32.60,23.14)(0.11,-0.22){12}{\line(0,-1){0.22}}
\multiput(33.96,20.48)(0.12,-0.41){7}{\line(0,-1){0.41}}
\multiput(34.77,17.60)(0.11,-1.49){2}{\line(0,-1){1.49}}
\multiput(35.00,14.63)(-0.09,-0.74){4}{\line(0,-1){0.74}}
\multiput(34.62,11.66)(-0.12,-0.35){8}{\line(0,-1){0.35}}
\multiput(33.67,8.83)(-0.11,-0.20){13}{\line(0,-1){0.20}}
\multiput(32.18,6.24)(-0.12,-0.13){17}{\line(0,-1){0.13}}
\multiput(30.20,4.00)(-0.15,-0.11){16}{\line(-1,0){0.15}}
\multiput(27.82,2.20)(-0.24,-0.12){11}{\line(-1,0){0.24}}
\multiput(25.13,0.90)(-0.41,-0.11){7}{\line(-1,0){0.41}}
\multiput(22.24,0.17)(-1.49,-0.07){2}{\line(-1,0){1.49}}
\multiput(19.25,0.02)(-0.74,0.11){4}{\line(-1,0){0.74}}
\multiput(16.30,0.46)(-0.31,0.11){9}{\line(-1,0){0.31}}
\multiput(13.49,1.49)(-0.20,0.12){13}{\line(-1,0){0.20}}
\multiput(10.94,3.04)(-0.13,0.12){17}{\line(-1,0){0.13}}
\multiput(8.75,5.07)(-0.12,0.16){15}{\line(0,1){0.16}}
\multiput(7.01,7.50)(-0.11,0.25){11}{\line(0,1){0.25}}
\multiput(5.78,10.22)(-0.11,0.49){6}{\line(0,1){0.49}}
\put(5.12,13.13){\line(0,1){2.99}}
\multiput(5.04,16.12)(0.10,0.59){5}{\line(0,1){0.59}}
\multiput(5.56,19.06)(0.11,0.28){10}{\line(0,1){0.28}}
\multiput(6.65,21.84)(0.12,0.18){14}{\line(0,1){0.18}}
\multiput(8.27,24.35)(0.12,0.12){18}{\line(0,1){0.12}}
\multiput(10.36,26.49)(0.16,0.11){15}{\line(1,0){0.16}}
\multiput(12.83,28.17)(0.28,0.12){10}{\line(1,0){0.28}}
\multiput(15.58,29.33)(0.74,0.11){6}{\line(1,0){0.74}}
\multiput(60.00,30.00)(0.99,-0.10){3}{\line(1,0){0.99}}
\multiput(62.97,29.70)(0.36,-0.11){8}{\line(1,0){0.36}}
\multiput(65.83,28.82)(0.22,-0.12){12}{\line(1,0){0.22}}
\multiput(68.45,27.39)(0.13,-0.11){17}{\line(1,0){0.13}}
\multiput(70.74,25.47)(0.12,-0.15){16}{\line(0,-1){0.15}}
\multiput(72.60,23.14)(0.11,-0.22){12}{\line(0,-1){0.22}}
\multiput(73.96,20.48)(0.12,-0.41){7}{\line(0,-1){0.41}}
\multiput(74.77,17.60)(0.11,-1.49){2}{\line(0,-1){1.49}}
\multiput(75.00,14.63)(-0.09,-0.74){4}{\line(0,-1){0.74}}
\multiput(74.62,11.66)(-0.12,-0.35){8}{\line(0,-1){0.35}}
\multiput(73.67,8.83)(-0.11,-0.20){13}{\line(0,-1){0.20}}
\multiput(72.18,6.24)(-0.12,-0.13){17}{\line(0,-1){0.13}}
\multiput(70.20,4.00)(-0.15,-0.11){16}{\line(-1,0){0.15}}
\multiput(67.82,2.20)(-0.24,-0.12){11}{\line(-1,0){0.24}}
\multiput(65.13,0.90)(-0.41,-0.11){7}{\line(-1,0){0.41}}
\multiput(62.24,0.17)(-1.49,-0.07){2}{\line(-1,0){1.49}}
\multiput(59.25,0.02)(-0.74,0.11){4}{\line(-1,0){0.74}}
\multiput(56.30,0.46)(-0.31,0.11){9}{\line(-1,0){0.31}}
\multiput(53.49,1.49)(-0.20,0.12){13}{\line(-1,0){0.20}}
\multiput(50.94,3.04)(-0.13,0.12){17}{\line(-1,0){0.13}}
\multiput(48.75,5.07)(-0.12,0.16){15}{\line(0,1){0.16}}
\multiput(47.01,7.50)(-0.11,0.25){11}{\line(0,1){0.25}}
\multiput(45.78,10.22)(-0.11,0.49){6}{\line(0,1){0.49}}
\put(45.12,13.13){\line(0,1){2.99}}
\multiput(45.04,16.12)(0.10,0.59){5}{\line(0,1){0.59}}
\multiput(45.56,19.06)(0.11,0.28){10}{\line(0,1){0.28}}
\multiput(46.65,21.84)(0.12,0.18){14}{\line(0,1){0.18}}
\multiput(48.27,24.35)(0.12,0.12){18}{\line(0,1){0.12}}
\multiput(50.36,26.49)(0.16,0.11){15}{\line(1,0){0.16}}
\multiput(52.83,28.17)(0.28,0.12){10}{\line(1,0){0.28}}
\multiput(55.58,29.33)(0.74,0.11){6}{\line(1,0){0.74}}
\put(55.00,23.00){\line(-1,0){30.00}}
\put(31.00,15.00){\line(1,0){18.00}}
\put(52.00,10.00){\line(-1,0){24.00}}
\put(28.00,10.00){\line(5,-1){24.00}}
\put(52.00,5.20){\line(-1,0){24.00}}
\put(28.00,5.20){\line(5,1){24.00}}
\put(15.00,15.00){\makebox(0,0)[cc]{{\Large L}}}
\put(65.00,15.00){\makebox(0,0)[cc]{{\Large R}}}
\put(24.00,23.00){\makebox(0,0)[rc]{$\bf s_1$}}
\put(56.00,23.00){\makebox(0,0)[lc]{$\bf s_{1'}$}}
\put(50.00,15.00){\makebox(0,0)[lc]{$\bf s_{2'}$}}
\put(30.00,15.00){\makebox(0,0)[rc]{$\bf s_2$}}
\put(27.00,10.00){\makebox(0,0)[rc]{$\bf s_3$}}
\put(53.00,10.00){\makebox(0,0)[lc]{$\bf s_{3'}$}}
\put(53.00,5.00){\makebox(0,0)[lc]{$\bf s_{4'}$}}
\put(27.00,5.00){\makebox(0,0)[rc]{$\bf s_4$}}
\end{picture}
\end{center}
\caption{Some possible crossing bonds.}
\end{figure}

Any state of the system can be expanded in terms of a square matrix
$c$,
\eq
\psi = \sum_{\al,\be} c_{\al \be}  \psi_\al \ot \widetilde \psi_\be ,
\label{psi}
\en
where $\{\psi_\al\}$ is a basis of $S^{(3)}$-eigenstates. (The
indices $\al$, $\be$ may contain additional non-spin 
quantum numbers, as needed, and the
tilde on the second tensor factor denotes the spin rotation.)
We shall assume that the state is normalized:
$\la \psi | \psi \ra = \tr c c^\dagger = 1$.
The energy expectation in terms of $c$ is a matrix expression
\eq
\la\psi|H|\psi\ra = \tr c c^\dagger h + \tr (c^\dagger c)^T h
- \sum_A \sum_{a=1}^3  \tr c^\dagger S_A^{(a)} c (S_A^{(a)})^\dagger ,
\label{ee}
\en
here $(h)_{\al\be} = \la\psi_\al|H_L|\psi_\be\ra 
= \la\widetilde\psi_\al|H_R|\widetilde\psi_\be\ra$,
$(S_A^{(a)})_{\al\be}  
=\la\psi_\al| \sum_{i \in A} j_i \,s^{(a)}_i  |\psi_\be\ra$,
and we have used $(\widetilde{S_A^{(a)}})^T = -(S_A^{(a)})^\dagger$.
(For $a = 1, 3$ the minus sign comes from the spin rotation, for $a=2$
it comes from complex conjugation. This can be seen by writing
$S^{(1)}$ and $S^{(2)}$ in terms of the real matrices $S^+$ and $S^-$.)
Note, that we do not assume $(h)_{\al\be}$ to be real or symmetric, otherwise
the following considerations would simplify considerably~\cite{LS}.

We see, by inspection, that the energy expectation value remains unchanged if
we replace $c$ by its transpose $c^T$, and, by linearity, if we replace it
by $c+c^T$ or $c-c^T$. So, if $c$ corresponds to a ground state, then we
might as well assume for convenience that $c$ is either symmetric or
antisymmetric. 
Note, that in either case we have $(c_R)^T = c_L$, where $c_L \equiv 
\sqrt{c c^\dagger}$ and $c_R \equiv \sqrt{c^\dagger c}$.
(Proof:
$(c_R^2)^T = (c^\dagger c)^T = c c^\dagger = c_L^2$, if $c^T = \pm c$;
now take the unique square root of this.) Using this we see that the 
first two terms in the energy
expectation equal $2 \tr c_L^2 h $ and thus depend on $c$ only through 
the positive semidefinite matrix $c_L$.
With the help of a trace inequality we will show that
the third term does not increase if we replace $c$ by the positive semidefinite
matrix $c_L$.

\section{Trace inequality}

For any square matrices $c$, $M$, $N$ it is true that \cite{KLS}
\eq
|\tr c^\dagger M c N^\dagger| \leq \frac{1}{2}\left( \tr c_L M c_L M^\dagger
+ \tr c_R N c_R N^\dagger \right) ,
\en
where $c_R = \sqrt{c^\dagger c}$, $c_L = \sqrt{c c^\dagger}$ are the
unique square roots of the positive semidefinite matrices $c^\dagger c$
and $c c^\dagger$. For the convenience of the reader we shall
repeat the proof here: By the polar decomposition theorem
$c = u c_R$ with a unitary matrix $u$ and
$(u c_R u^\dagger)^2 = u c^\dagger c u^\dagger = c c^\dagger = c_L^2$, 
so by the uniqueness of the square root
$u c_R u^\dagger = c_L$. Similarly,
for any function $f$ on the non-negative
real line $u f(c_R) u^\dagger$ = $f(c_L)$, and in particular
$u \sqrt{c_R} = \sqrt{c_L} u$
and thus $c = \sqrt{c_L} u \sqrt{c_R}$.
Let $P \equiv u^\dagger \sqrt{c_L} M \sqrt{c_L} u$ and
$Q \equiv \sqrt{c_R} N^\dagger \sqrt{c_R}$, then
\eqa
|\tr c^\dagger M c N^\dagger| & = & |\tr P Q| 
\, \leq \, \frac{1}{2}(\tr P P^\dagger + \tr Q Q^\dagger) \nn
& = & \frac{1}{2}\left( \tr c_L M c_L M^\dagger
+ \tr c_R N c_R N^\dagger \right) ,
\ena
where the inequality is simply the geometric arithmetic
mean inequality for matrices
$$
|\tr P Q| = |\sum_{i,j} P_{ij} Q_{ji}| \leq 
\frac{1}{2} \sum_{i,j} (|P_{ij}|^2 +  |Q_{ji}|^2)
= \frac{1}{2}(\tr P P^\dagger + \tr Q Q^\dagger) .
$$

\section{Existence of a positive ground state}
\label{positive}

Consider any ground state of the system with coefficient matrix 
$c = \pm c^T$
and apply the trace inequality to the terms in $\la\psi|H_C|\psi\ra$:
$$
-\tr c^\dagger S_A^{(a)} c (S_A^{(a)})^\dagger
\geq
-\frac{1}{2} \left(\tr c_L S_A^{(a)} c_L (S_A^{(a)})^\dagger
+ \tr c_R S_A^{(a)} c_R (S_A^{(a)})^\dagger\right) ,
$$
but $(c_R S_A^{(a)} c_R (S_A^{(a)})^\dagger)^T 
= ((S_A^{(a)})^T)^\dagger c_L (S_A^{(a)})^T c_L
= (S_A^{(a)})^\dagger c_L (S_A^{(a)}) c_L$, so in fact
$$
-\tr c^\dagger S_A^{(a)} c (S_A^{(a)})^\dagger 
\geq -\tr c_L S_A^{(a)} c_L (S_A^{(a)})^\dagger .
$$
Since the normalization of the state and 
the other terms in (\ref{ee}) are unchanged if we replace $c$ by 
$c_L = \sqrt{c c^\dagger}$, and because we have assumed that $c$ is 
the coefficient matrix of a ground state, it follows that the 
positive semidefinite matrix $c_L$ must also be the coefficient 
matrix of a ground state.

\section{Overlap with canonical spin zero state}

Consider the (not normalized) canonical state with coefficient matrix given by
the identity matrix in a basis of $S^{(3)}$-eigenstates of either
subsystem
\eqa
\Xi & = & \displaystyle \sum_{k,k'}   
\sum_{j}\sum_{m = -j}^j \psi_{(j,m,k)} 
\ot \widetilde \psi_{(j,m, k')} \nn
& = & \displaystyle \sum_{k,k'}       
\sum_{j}\sum_{\tilde m = -j}^j \psi_{(j,m,k)} 
\ot(-)^{j-m} \psi_{(j,-m, k')}.
\ena
The states are labeled by the usual spin quantum numbers $j$, $m$ and an 
additional symbolic quantum number $k$ to lift remaining ambiguities. 
The state $\Xi$ has total spin zero because of the spin rotation in 
the right subsystem:
Its $S_{\text{tot}}^{(3)}$-eigenvalue is zero and acting with either 
$S^+_{\text{tot}}$ or $S^-_{\text{tot}}$ on it gives zero.
The overlap of any state with coefficient matrix $c$ with the canonical
state $\Xi$ is simply the trace of $c$. 
In the previous section we found that the reflection symmetric spin
system necessarily has a ground state with positive semidefinite, non-zero 
coefficient matrix, which, by definition, has a (non-zero) positive trace. 
Since the trace is proportional to the overlap with the canonical spin-zero
state, we have now shown that there is always a ground state that
contains a spin-zero part. Provided that total spin is a good quantum
number, we can conclude further that our system always has a ground
state with total spin zero, i.e., a singlet. 

\section{Projection onto spin zero}

Consider any state
$\psi = \sum c_{\al \be} \psi_\al \ot \widetilde \psi_\be$
with positive semidefinite $c = |c|$. We have seen that this implies
that $\psi$ has a spin-zero component. If total spin is a good quantum
number it is interesting to ask what happens to $c$ when we project
$\psi$ onto its spin zero part 
\eq
\psi^0 = \sum c^0_{\al \be} \psi_\al \ot \widetilde \psi_\be .
\en
We shall show that the
coefficient matrix $c^0$ of $\psi^0$ is a partial trace of $c$ and thus
still positive semidefinite.  A convenient parametrisation of the 
$S^3$ eigenstates $\psi_\al$ for this task is, as before, 
$\al = (j,m,k)$, where $k$ labels spin-$j$
multiplets $[j]_k$ in the decomposition of the Hilbert space
of one subsystem into components of total spin.
Note that $[j]_k \ot [j']_{k'} = [j+j']  \op \ldots \op [|j-j'|]$, so
$[j]_k \ot [j']_{k'}$ contains a spin zero subspace only if $j = j'$, and for
each $k$, $k'$ that subspace is unique and generated by the normalized
spin zero state
\eq
\xi_{k,k'} = (2j+1)^{-\frac{1}{2}} 
\sum_{\tilde m = -j}^j \psi_{(j,\tilde m,k)} 
\ot \widetilde \psi_{(j,\tilde m, k')} .
\en
(Recall that
$\widetilde \psi_{(j,\tilde m, k')} 
= (-)^{j-\tilde m} \psi_{(j,-\tilde m, k')}$
is the rotation of $\psi_{(j,\tilde m, k')}$ by $\pi$ around the 2-axis in
spin space.)
The projection of $\psi$ onto spin zero is thus amounts to replacing $c$ with
$c^0$, where
\eq
c^0_{(j,m,k)(j',m',k')} = \left\{
\begin{array}{l}
0 \quad\mbox{if $j \neq j'$ or $m \neq m'$} \\
\displaystyle \frac{N}{2j + 1} \sum_{\tilde m = -j}^j 
c_{(j,\tilde m, k)(j,\tilde m, k')} \quad\mbox{else} .
\end{array}
\right. \label{czero}
\en
($N$ is a overall normalisation constant, independent of $j$, $m$, $k$.)
Let us now show that this partial trace preserves positivity, i.e.,
$(v, c^0 \; v) \geq 0$ for any vector $v=\left(v_{(j,m,k)}\right)$ 
of complex numbers.
If we decompose $v$ into a sum of vectors $v_{j m}$
with definite $j$, $m$ and use (\ref{czero}), we see
\eq
(v , c^0 \, v)  =  \sum_{j,m} (v_{j m} , c^0 \, v_{j m})
=  \sum_{j,m,\tilde m} (\omega_{j \tilde m}^m ,c \, \omega_{j \tilde m}^m)
\geq 0 ,
\en
where the $\omega_{j \tilde m}^m$ are new vectors with components
$\omega_{(j,\tilde m,k)}^m = v_{(j,m,k)}$, independent of $\tilde m$. 
Every term in the last sum
is non-negative because $c$ is positive semidefinite by assumption.
This result implies in particular that a reflection symmetric spin
system always has a ground state with total spin zero \emph{and} positive
semidefinite coefficient matrix -- provided that total spin is a good
quantum number.

\section{Ice rule for crossing bonds}

The expectation of the third spin component
of the sites involved in each crossing bond $B$, weighted by their coefficients
$j_i$, vanishes for \emph{any} ground state $\psi_0$,
\eq
\la \psi_0 |\sum_{i \in B} j_i (s_i^{(3)} + s_{i'}^{(3)})|
\psi_0\ra = 0,
\label{ice}
\en
provided that either the left
and right subsystems are invariant under the spin rotation,
$h = \tilde h$, or that their matrix elements are real (the latter
is equivalent to the assumption $h = h^T$, since we know
that $h = h^\dagger$ or otherwise the whole spin Hamiltonian would not be 
Hermitean).
By symmetry (\ref{ice}) will also be true for the first spin component and, if 
we are dealing with a spin Hamiltonian that is invariant under spin rotations, 
it is also true for the second spin component.
For ground states with symmetric or antisymmetric coefficient matrix
we automatically have $\la s_i^{(3)} + s_{i'}^{(3)} \ra = 0$ for
any pair of sites $i$ and $i'$, so in that case (\ref{ice}) is trivial.

For the proof we introduce a real parameter $b$ in the spin Hamiltonian:
$H(b) \equiv H - b(S^{(3)}_B + S^{(3)}_{B'}) + b^2/2$, where $B$ is one
of the sets of sites involved in the crossing bonds of the original
Hamiltonian $H$. Let $E_b$ be the 
ground state energy of $H(b)$ and $E_0$ the ground state energy of $H$.
One can show that $E_b \geq E_0$ and
(\ref{ice}) follows then by a variational argument:
\eq
\la \psi_0 | H(b) | \psi_0 \ra \geq E_b \geq E_0 = \la \psi_0 | H | \psi_0 \ra,
\en
or, $\la\psi_0| b(S^{(3)}_B + S^{(3)}_{B'})|\psi_0\ra \leq b^2/2$, which
implies (\ref{ice}). Note, that we did not make any assumptions about the symmetry
or antisymmetry of the coefficient matrix of $\psi_0$ here.

Sketch of the proof of $E_b \geq E_0$ (see also \cite{KLS,LS}): 
$H(b) = H_L(b) + H_R(b) + H_C(b) + b^2/4$ with 
$H_{L,R}(b) = H_{L,R} - b/2 \cdot S^{(3)}_B$ and 
$H_C(b)$ equal to $H_C$ except for the term $S^{(3)}_B \cdot S^{(3)}_{B'}$, which
is replaced by $(S^{(3)}_B - b/2)\cdot(S^{(3)}_{B'} - b/2)$.
If we now write the ground state energy expectation of $H(b)$ as 
a matrix expression like (\ref{ee}) and apply the trace inequality to
it, we will find an equal or lower energy expectation
not of $H(b)$, but rather of $H$: The trace inequality effectively removes the 
parameter $b$ from the Hamiltonian. By the variational principle the true
ground state energy of $H$ is even lower and we
conclude that $E_b \geq E_0$. Role of the technical assumptions mentioned above:
If $h = h^T$, then the transpose in the second term in (\ref{ee}) vanishes,
the matrix expression is symmetric in $c_L$ and $c_R$ (except for the sign
of the parameter $b$), and the trace inequality gives
$\la H(b) \ra_c \geq \frac{1}{2}\left\{\la H \ra_{c_L} + \la H \ra_{c_R}\right\}$.
If $h = \tilde h$, then we should drop the spin rotation  on the
second term of the analog of expression (\ref{psi}) for $\psi_b$. 
The matrix expression for $\la H(b) \ra$ is then symmetric in $c$ and $c^T$ 
and we may assume $c = \pm c^T$ to prove $E_b \geq E_0$. The calculation is
similar to the one in section~\ref{positive}. Note, that $c = \pm c^T$ only
enters the proof of $E_b \geq E_0$, we still do not need to assume that the coefficient
matrix of $\psi_0$ in (\ref{ice}) has that property.

The preferred configurations of
four spins with antiferromagnetic crossed bonds in a classical
Ising system are very similar to the configurations of the 
four hydrogen atoms that surround each oxygen
atom in ice: There are always two hydrogen atoms close and two further
away from each oxygen atom, and there are always
two spins ``up'' and two ``down'', i.e. $M = 0$, in the Ising system.
Equation $(\ref{ice})$
is a (generalized) quantum mechanical version of this --
that is why we use the term ``ice rule''. This phrase
is also used in the context of \emph{ferromagnetic} pyrochlore
with Ising anisotropy (``spin ice'') \cite{HBMZG} and we hope that does not cause
confusion.

\section{Discussion}

We would like to discuss similarities between our method and 
previous work, in particular the approach of \cite{LM} for
the bipartite antiferromagnet:
There, the spin Hamiltonian splits into two parts $H = H_0 + H_1$. 
The expectation value of $H_0$ with respect to a state
$\psi = \sum f_\al \phi_\al$, expanded in an appropriate basis
$\{\phi_\al\}$, depends only on $|f_\al|$, 
and the expectation of $H_1$ does not increase under the transformation 
$f_\al \rightarrow |f_\al|$. The variational principle then implies that there must be 
a ground state with only non-negative coefficients $|f_\al|$.
The present setup is very similar, except that 
we use coefficient matrices $(c_{\al \be})$ to expand states,
since we work on a tensor product of Hilbert spaces. In our case
the expectation value of
$H_0 = H_L + H_R$ depends only on $c$ via the positive matrices
$c_L$ and $c_R$, and the expectation value of $H_C$ increases
if we ``replace" $c$ by these positive matrices. The similarity is even
more apparent if $h$ has real matrix elements: In that case we may
assume that $c$ is diagonalisable and its eigenvalues play the
role of the coefficients $f_\al$. 
The spin of a positive ground state is established in all cases
from the overlap with a state of known spin that is also positive.
In a system with sufficient symmetry we can, however, also use the
``ice rule'' to prove that \emph{all} ground states have
total spin zero \cite{LS}. (E.g., in a system with constant coefficients
$j_i$ and enough translational invariance, so that every spin can
be considered to be involved in a crossing bond and thus in an ice rule,
we would conclude that all ground states have $S^{(3)}_{\text{tot}} = 0$
and, assuming rotational invariance in spin space,
$S_{\text{tot}} = 0$.)
It is not clear, if $M$-subspace methods can be used in the present setting
to get information about excited states.
An important point in the our work is that we 
consider not only antiferromagnetic
bonds between single sites but also bonds between sets of sites. This frees us
from the requirement of bipartiteness and even allows some ferromagnetic
crossing bonds, for example in $(s_1 - s_2)(s_{1'} - s_{2'})$.
There is no doubt that the scheme can be further generalized, e.g., to other
groups or more abstract ``crossing bonds''. In the present form the most interesting
applications are in the field of frustrated spin systems \cite{LS}.

We did not address the question of the degeneracy of ground states. Classically
a characteristic feature of frustrated systems is 
their large ground state degeneracy. For frustrated quantum spin systems this is an
important open problem.\\[1ex]
We would like to thank Roderich Moessner for inspiring discussions.

\end{document}